\documentclass[prl,amssymb,amsmath,reprint]{revtex4-1}
\usepackage{bm}
\usepackage{graphicx}
\usepackage{color}
\newcommand{\ve}{\mathbf}
\usepackage{bbm}
\begin{document}

\title{\normalsize Diagrammatic Approach for the High-Temperature Regime of Quantum Hall Transitions
}

\author{Martina Fl\"{o}ser}
\affiliation{Institut N\'{e}el, CNRS and Universit\'{e} Joseph Fourier, B.P.
166, 25 Avenue des Martyrs, 38042 Grenoble Cedex 9, France}

\author{Serge Florens}
\affiliation{Institut N\'{e}el, CNRS and Universit\'{e} Joseph Fourier, B.P.
166, 25 Avenue des Martyrs, 38042 Grenoble Cedex 9, France}

\author{Thierry Champel}
\affiliation{Universit\'e Joseph Fourier Grenoble I / CNRS UMR 5493,
Laboratoire de Physique et Mod\'elisation des Milieux Condens\'es, B.P. 166, 
38042 Grenoble, France}

\begin{abstract}
We use a general diagrammatic formalism based on a local conductivity approach
to compute electronic transport in continuous media with long-range disorder, in the 
absence of quantum interference effects.
The method allows us then to investigate the interplay of dissipative processes and random 
drifting of electronic trajectories in the high-temperature regime of quantum Hall 
transitions. 
We obtain that the longitudinal conductance $\sigma_{xx}$ scales with an exponent 
$\kappa=0.767\pm0.002$ in agreement with the value $\kappa=10/13$ conjectured from 
analogies to classical percolation. We also derive a microscopic expression for the 
temperature-dependent peak value of $\sigma_{xx}$, useful to extract $\kappa$
from experiments.
\end{abstract}


\maketitle

{\it Introduction.--} The geometric concept of percolation is ubiquitous to electronic
transport in strongly disordered media~\cite{Isichenko}, in both the classical and quantum realm.
Indeed, building on earlier studies in the context of metallic
alloys and granular materials~\cite{Kirkpatrick1973}, recent advances have 
extended percolation ideas to the description of quantum phases in low-dimensional electron 
gases, ranging from metal/insulator transitions at low magnetic 
field to the high magnetic field regime associated to the quantum Hall 
effect~\cite{Meir1999,Kettemann,EversRMP}.
Despite this very seductive geometrical analogy, difficulties arise 
for a microscopic description of transport because the electrical current does not 
just propagate on simple geometrical objects, such as the bulk or the boundaries
of a percolation network. In fact, in a dissipative system the current density 
always spreads along extended structures, so that fractality of the transport 
network may be smeared in realistic situations~\cite{SH1994}.
While fully numerical simulations of transport models can account
for such complexity~\cite{Meir1999,EversRMP}, they bring finite size effects
that give limitations for quantitative description of transport.
For instance, an important question for metrological purposes~\cite{Cage2005} is the 
precise understanding of the accuracy of Hall conductance quantization, where percolation 
is known to play a role, both from theoretical grounds~\cite{SH1994,PS1995,FS1995} and 
from local density of states~\cite{HSWIM2008} and transport 
measurements~\cite{Renard2004,Zhao2008,Li2010}.

\begin{figure}[ht]
\hspace{1cm}\includegraphics[scale=0.7]{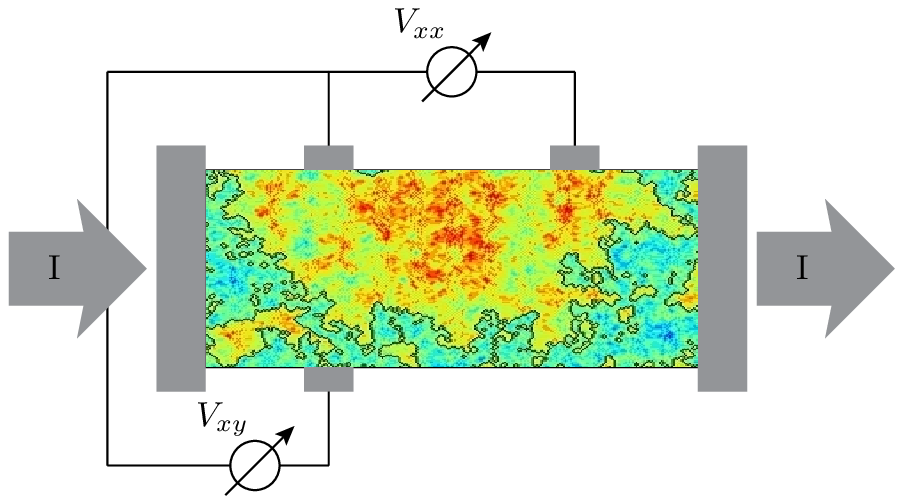}
\caption{(color online) Two-dimensional sample with percolating random charge
inhomogeneities: measurement of longitudinal $V_{xx}$ and Hall $V_{xy}$ voltages with
applied current $I$.}
\label{sample}
\end{figure}

Our goal in this Letter is to show that percolation features of transport in continuous 
disordered media can be captured analytically by a diagrammatic approach, starting from local 
Ohm's law: 
\begin{align} 
\ve{j}(\ve{r})=\hat\sigma(\ve{r})\ve{E}(\ve{r}),
\end{align} 
with $\ve{j}$ the local current density and $\ve{E}$ the local electrical field.
This introduces $\hat\sigma(\ve{r})$ the local conductivity tensor, a spatially-dependent quantity
due to inhomogeneities, that naturally encodes altogether dissipation, disorder and
confinement~\cite{SH1994,ICS2006,PP2007}. The local conductivity model is expected to be accurate 
at high enough temperatures whenever phase-breaking processes, such as electron-phonon 
scattering, occur on length scales that are shorter than the typical variations of 
disorder. However, quantum mechanics may still be important to determine microscopically 
the quantitative behavior of the local conductivity tensor~\cite{GV1994,CFC2008}. 
The main difficulty thus lies in solving 
the continuity equation $\bm\nabla\cdot\ve{j}=0$ in the presence of long-range random 
inhomogeneities in the sample, see Fig.~\ref{sample}.
%
%


{\it General formalism.--}
Our starting point follows early ideas proposed by several
authors~\cite{Dreizin1972,Stroud1975},  
where effective conductivity approaches were developed based
on a local conductivity tensor $\hat\sigma(\ve{r})$.
We consider the general situation of an arbitrary and continuous distribution of 
conductivity in a macroscopic $d$-dimensional sample of volume $V$, bounded by a 
surface $S$. 
The experimentally accessible quantity is the average current density 
$\langle \ve{j}\rangle=\hat\sigma_\mathrm{eff}\ve{E}_0$ which is driven by applying a constant electric 
field $\ve{E}_0$ at the boundary of the sample. This defines a position-independent 
effective conductivity tensor $\hat\sigma_\mathrm{eff}$, which is nothing but the 
macroscopic conductance tensor, up to a geometrical prefactor. 
Following Ref.~\onlinecite{Stroud1975}, we decompose (arbitrarily at this stage) 
$\hat\sigma(\ve{r})=\hat\sigma_0+\delta\hat\sigma(\ve{r})$ into uniform and 
fluctuating parts, respectively.
By expressing the electrical field by its scalar potential
$\ve{E}(\ve{r})=-\bm\nabla\Phi(\ve{r})$, the continuity equation leads
to the boundary value problem:
\begin{align}
\bm\nabla\cdot\left[\hat\sigma_0\bm\nabla\Phi(\ve{r})\right]
=-\bm\nabla\cdot\left[\delta\hat\sigma(\ve{r})\bm\nabla\Phi(\ve{r})\right]\text{ in V}\\
\Phi(\ve{r})\equiv\Phi_0(\ve{r})=-\ve{E}_0\cdot\ve{r}\text{ on S.}
\end{align}
By introducing the Green's function $G(\ve{r},\ve{r'})$ defined by
\begin{align}
\bm\nabla\cdot[\hat\sigma_0\bm\nabla G(\ve{r},\ve{r'})]=-\delta(\ve{r}-\ve{r'}) 
\text{ in V}\label{TAT:eq:Gboundary}\\
G(\ve{r},\ve{r'})=0 \text{ for $\ve{r}$ on S,}\label{TAT:eq:Gboundary2}
\end{align}
the scalar potential is formally given by
\begin{align}
\Phi(\ve{r})=\Phi_0(\ve{r})
+\int_V\!\!\!\! d^dr'\; G(\ve{r},\ve{r'})
\bm\nabla'\cdot[\delta\hat\sigma(\ve{r'})\bm\nabla'\Phi(\ve{r'})]
\label{eq:Phi}
\end{align}
with the short-hand notation $\bm\nabla'=\bm\nabla_{\ve{r}'}$.
Integrating by parts with $\bm\nabla' G(\ve{r},\ve{r'})=-\bm\nabla G(\ve{r},\ve{r'})$ and taking 
the gradient on both sides of Eq.~(\ref{eq:Phi}) leads to
\begin{align}
\ve{E}(\ve{r})&=\ve{E}_0+\int_V\!\!\!\! d^dr'\; \bm\nabla\cdot\left[\bm\nabla G(\ve{r},\ve{r'})
\delta\hat\sigma(\ve{r'})\ve{E}(\ve{r'})\right]\\
&=\ve{E}_0+\int_V\!\!\!\! d^dr'\; \hat{\cal{G}}_0(\ve{r},\ve{r'})
\delta\hat\sigma(\ve{r'})\ve{E}(\ve{r'}),
\label{TAT:eq:E}
\end{align}
where
$\left[\hat{\cal{G}}_0\right]_{ij}=\frac{\partial}{\partial
r_i}\frac{\partial}{\partial r_j}G(\ve{r},\ve{r'})$.
Finally, multiplying Eq.~(\ref{TAT:eq:E}) by $\delta\hat\sigma(\ve{r})$ and
introducing a new local tensor $\hat\chi$ such that
$\delta\hat\sigma(\ve{r})\ve{E}(\ve{r})=\hat\chi(\ve{r})\ve{E}_0$,
we obtain:
\begin{align}\label{TAT:eq:chiE}
 \hat\chi(\ve{r})\ve{E}_0=\delta\hat\sigma(\ve{r})\ve{E}_0
+\delta\hat\sigma(\ve{r})\int_V\!\!\!\! d^dr'\;
\hat{\cal{G}}_0(\ve{r},\ve{r'})\hat\chi(\ve{r'})\ve{E}_0 .
\end{align}
As Eq.~\eqref{TAT:eq:chiE} is valid for all possible choices of $\ve{E}_0$, the 
following tensorial equation also holds:
\begin{align}\label{TAT:eq:chifinal}
 \hat\chi(\ve{r})=\delta\hat\sigma(\ve{r})
+\delta\hat\sigma(\ve{r})\int_V\!\!\!\! d^dr'\;
\hat{\cal{G}}_0(\ve{r},\ve{r'})\hat\chi(\ve{r'}) .
\end{align}
Spatial averaging of the current
$\ve{j}(\ve{r})=[\hat\sigma_0+\hat\chi(\ve{r})]\ve{E}_0$ over conductivity fluctuations 
$\delta\hat\sigma(\ve{r})$ leads therefore to the effective conductivity
 $\hat\sigma_\mathrm{eff}=\hat\sigma_0+\langle\hat\chi\rangle$,
where the spatial average on $\hat\chi$ is performed while enforcing the integral
equation~(\ref{TAT:eq:chifinal}). Although sample boundaries could be considered
in principle, 
 we now focus on an infinite 
sample, so that the Green's function [Eq. (\ref{TAT:eq:Gboundary})] becomes 
translation-invariant
\begin{align}
G(\ve{r},\ve{r}')=\int\!\!\!\frac{d^dp}{(2\pi)^d}\frac{e^{i\ve{p}\cdot(\ve{r}-\ve{r}')}}{\ve{p}\hat\sigma_0\ve{p}+0^+},
\label{eq:Green}
\end{align}
where $0^+$ is a small positive quantity which ensures the correct boundary
condition at infinity [Eq.~\eqref{TAT:eq:Gboundary2}].

\begin{figure}[ht]
\includegraphics{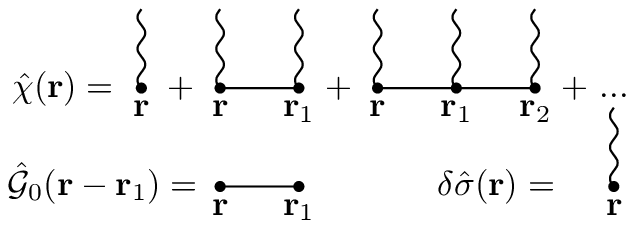}\\
\caption{Graphical representation of the strong-dissipation
expansion~(\ref{eq:ChiExpand}) of the self-consistent transport
equation~(\ref{TAT:eq:chifinal}).}
\label{fig:self}
\end{figure}

{\it Systematic expansion at strong-dissipation.--} 
Previous works either considered a mean-field solution of
Eq.~(\ref{TAT:eq:chifinal}) in the peculiar case of
binary randomness in the local conductivity tensor~\cite{Stroud1975}, or
computed low order contributions for continuous disorder distribution~\cite{Dreizin1972,TRO2005}. 
Our aim is to present a systematic expansion controlled by weak
fluctuations of the conductivity and to show that the nonperturbative regime
of large conductivity fluctuations can be tackled by sufficient knowledge of
the perturbative series.
The spatial average on $\hat\chi(\ve{r})$ can be obtained clearly after iterating
Eq.~(\ref{TAT:eq:chifinal}) to all orders:
\begin{align}
\label{eq:ChiExpand}
\langle \hat\chi(\ve{r})\rangle=&\langle\delta\hat{\sigma}(\ve{r})\rangle
+\int\!\! d^d\ve{r}_1 
\langle\delta\hat{\sigma}(\ve{r})\hat{\cal{G}}_0(\ve{r},\ve{r}_1)\delta\hat\sigma(\ve{r}_1)\rangle\\
\nonumber
&\hspace{-1.5cm}+\int\!\! d^d\ve{r}_1\!\!\int\!\! d^d\ve{r}_2 \langle\delta\hat{\sigma}(\ve{r})\hat{\cal{G}}_0(\ve{r},\ve{r}_1)
\delta\hat\sigma(\ve{r}_1)\hat{\cal{G}}_0(\ve{r}_1,\ve{r}_2)\delta\hat\sigma(\ve{r}_2)\rangle+...
\end{align}
which can be expressed graphically as in Fig.~\ref{fig:self}.
For incoherent transport, self-averaging occurs and the spatial average over 
the local conductivity fluctuations may be replaced by an ensemble average. 

Let us first illustrate the method for a purely resistive and isotropic medium, so that
$\hat\sigma_0= \sigma_0 \hat1$ and $\delta\hat{\sigma}(\ve{r})=
\delta\sigma(\ve{r})\hat1$, with $\left<\delta\sigma(\ve{r})\right>=0$.
In the limit of strong-dissipation compared to the typical fluctuations of conductivity [$\sigma_0\gg\sqrt{\langle\delta\sigma^2\rangle}$], we get
$\hat\sigma_\mathrm{eff}=\sigma_{xx}\hat1$ with
\begin{eqnarray}
\label{eq:fluctuation}
\sigma_{xx}&=&\sigma_0-\frac{1}{\sigma_0}\int\!\! d^dr 
\!\! \int\!\!\frac{d^dp}{(2\pi)^d}\frac{p_x^2e^{i\ve{p}\ve{r}}}
{\ve{p}^2+0^+}\langle\delta\sigma(\ve{r})\delta\sigma(\ve{0})\rangle\\
\nonumber
&=&\sigma_0-\frac{1}{\sigma_0}\int\!\! d^dr \;\frac{\delta(\ve{r})}{d}
\langle\delta\sigma(\ve{r})\delta\sigma(\ve{0})\rangle
=\sigma_0-\frac{\langle\delta{\sigma}^2\rangle}{d\sigma_0}.
\end{eqnarray}
We thus recover previous results~\cite{TRO2005} obtained for weakly
disordered media, which predict a reduction of the macroscopic conductance 
due to randomly distributed resistive barriers. Clearly, nontrivial geometrical 
aspects are absent at this order, because the dominant background of
conductivity $\sigma_0$ prevents the percolating network to establish.
This general formulation of transport [Eq.~(\ref{eq:ChiExpand})] is
immediately appealing because arbitrary orders of the strong-dissipation expansion
can be generated in a compact fashion, fostering hope that the difficult limit
of large conductivity fluctuations can be tackled by standard resummation methods.

\begin{figure}
\includegraphics[width=0.9\linewidth]{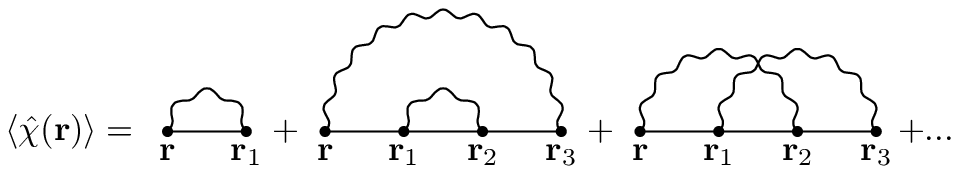}
\caption{Diagrammatic expansion in the case of Gaussian
fluctuations of the local conductivity. Wiggly lines denote the conductivity
correlation functions.
}
\label{fig:diag}
\end{figure}

{\it Simplification for Gaussian randomness.--}
Under some microscopic assumptions, the conductivity tensor may 
follow a random Gaussian distribution, according to 
$\left<\delta\hat{\sigma}(\ve{r})\right>=0$ and
$\left<\delta{\sigma}_{ij}(\ve{r}) \delta{\sigma}_{kl}(\ve{r'})\right>
=C_{ij;kl}(\ve{r}-\ve{r'})$, so that all moments of the local
conductivity tensor are determined from Wick's theorem (in particular, all odd
correlations vanish here). This hypothesis leads to a familiar-looking
diagrammatic formulation for the strong-dissipation expansion, as shown in
Fig.~\ref{fig:diag}.
An important technical point is that all particle reducible graphs (diagrams
that can be split in two parts by cutting a single line of $\hat{\cal{G}}_0$) 
are identically zero. This is because all such contributions contain the
zero momentum limit of the Green's function $[\hat{\cal{G}}_0]_{ij}(\ve{p})
=-p_i p_j G(\ve{p})$ which vanishes at zero momentum
according to Eq.~(\ref{eq:Green}) (note the crucial role of the regularization
parameter). Interestingly, the conductance correction $\left<\hat\chi\right>$
now takes the precise form of a self-energy, in contrast to a fully quantum formulation
of electronic transport~\cite{Montambaux} where vertex corrections associated to 
interference effects need to be accounted for. In what follows, we wish to use the method with the challenging regime
of a strongly fluctuating local conductivity, that may lead to geometrical effects 
related to classical percolation. Clearly, the general perturbation
series~(\ref{eq:ChiExpand}) in powers of $\langle\delta{\hat\sigma}^2\rangle/\sigma_0^2$ 
then breaks down, so that high order terms will be needed.

{\it Percolation regime of the semiclassical Hall effect.--}
We 
henceforth  consider 
the semiclassical regime of
the quantum Hall effect, which occurs in very high mobility two-dimensional electron
gases at large perpendicular magnetic field~\cite{Zhao2008,Li2010}.
General physical arguments~\cite{SH1994} as well as microscopic
calculations~\cite{GV1994,CFC2008} show that the electron dynamics 
can be described in this regime by a local Ohm's law with a randomly fluctuating 
Hall conductivity $\sigma_H(\ve{r})=\sigma_H+\delta\sigma(\ve{r})$:
\begin{equation}
\label{TAT:eq:condTensor}
\hat\sigma(\ve{r})=
\begin{pmatrix}
\hspace{-1.35cm}\sigma_0 &\hspace{-1cm} -\sigma_H-\delta\sigma(\ve{r})\\
\sigma_H+\delta\sigma(\ve{r}) & \;\;\;\;\;\sigma_0
\end{pmatrix}
\;\mathrm{with}\; \langle\delta\sigma(\ve{r})\rangle=0.
\end{equation}
According to the classical Hall's law, such purely off-diagonal fluctuations of the 
conductivity correspond to spatial modulations of the electron density brought by 
long-range random impurities~\cite{SH1994,ICS2006}.
The diagonal part in Eq.~(\ref{TAT:eq:condTensor}) accounts phenomenologically
for dissipative processes, such as electron-phonon scattering, and is supposed
for simplicity to be spatially uniform. 

The explicit connection to geometrical percolation can now be made.
At vanishing dissipation $\sigma_0\to0$, drift currents follow
from Hall's law and propagate along constant lines of Hall conductivity.
Indeed, from Maxwell's equation $\bm\nabla\times\ve{E}=0$ and current 
conservation $\bm\nabla\cdot\ve{j}=0$, one gets the transport equation
$\left[\bm\nabla\delta\sigma(\ve{r})\right]\cdot\ve{j}=0$. The lines of 
constant $\delta\sigma(\ve{r})$ are typically closed, so that all electronic states 
are localized, except the ones living on the percolation cluster. However,
the percolating state does not contribute to macroscopic transport either, as
it must necessarily pass through saddle-points of the disordered landscape, where
the transport equation becomes undetermined. Thus having finite 
$\sigma_0$ is required to establish a finite conductance in the sample, by
connecting the different nearly localized states. This difficulty has led 
authors~\cite{SH1994} to wonder whether purely geometric arguments are sufficient 
to understand the transport properties at small but finite dissipation, because
the current carrying states become broad filaments that may smear the fractal
structure of the percolation cluster. This question is now investigated 
in a controlled fashion.

At high temperature, the Hall conductivity fluctuations 
given by Eq.~(\ref{TAT:eq:condTensor}) follow the Gaussian 
distribution of disorder~\cite{Supplementary}. 
We also consider for simplicity Gaussian spatial
correlations $ \langle\delta\sigma(\ve{r})\delta\sigma(\ve{r'})\rangle=
\langle\delta\sigma^2\rangle e^{-|\ve{r}-\ve{r'}|^2/\xi^2}$,
with correlation length $\xi$.
Inspection of the diagrammatic series depicted in Fig.~\ref{fig:diag} shows
that the effective conductivity obeys the following expansion:
\begin{equation}
 \hat\sigma_\mathrm{eff}=\begin{pmatrix}0 \! &-\sigma_H\\\sigma_H \! &0\end{pmatrix}
+\sigma_0\left[1+\sum_{n=1}^\infty a_{n}\frac{\langle\delta\sigma^2\rangle^n}{\sigma_0^{2n}}\right]
\begin{pmatrix}1&0\\0&1\end{pmatrix}
\label{eq:series}
\end{equation}
with dimensionless coefficients $a_n$ collecting all diagrams of order $n$ in
perturbation theory in $\langle\delta\sigma^2\rangle/\sigma_0^2$.
The Hall component is therefore not affected here, while the longitudinal
conductance receives nontrivial corrections that encode the interplay of
dissipation and percolation.
The diagrammatic formulation of transport allowed us to compute 
this series up to sixth order~\cite{Supplementary}.

\begin{figure}
 \includegraphics[width=0.8\linewidth]{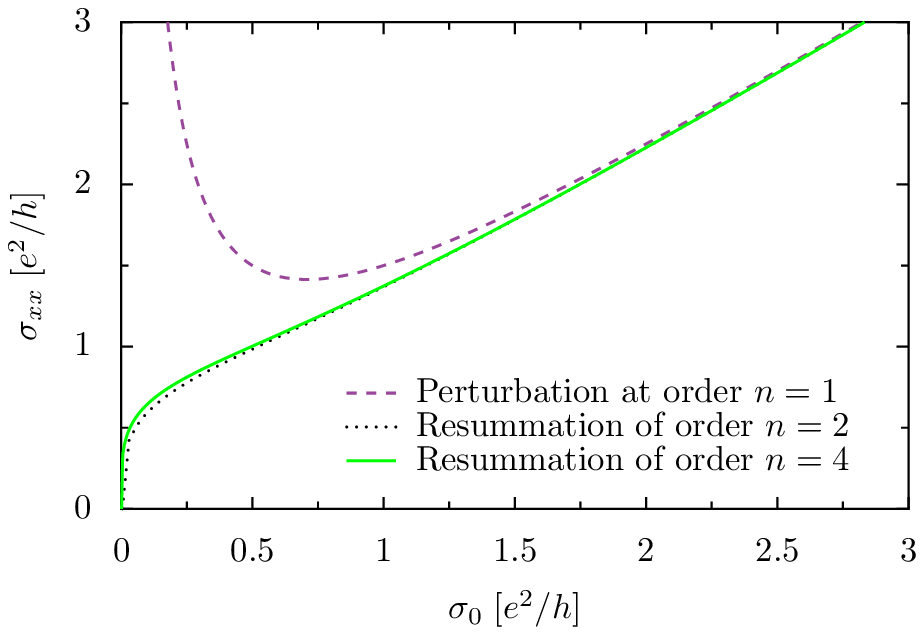}
 \caption{\label{TAT:fig:cond} (color online) Scaling function of the
longitudinal
conductance from the percolating ($\sigma_0\to0$) to the dissipative
regime ($\sigma_0\to\infty$). A comparison is made between first-order
bare perturbation theory (controlled only at large dissipation) to the
resummations of
the $n=2$ and $n=4$ orders, showing good convergence for all values of
$\sigma_0$.}
\label{fig:extrapolation}
\end{figure}

As understood previously, the effective longitudinal conductivity $\sigma_{xx}$ must 
vanish when $\sigma_0\to0$ for a continuous local conductivity model,
and previous works~\cite{SH1994,PS1995,FS1995} suggested 
a power-law dependence $\sigma_{xx}\sim C \langle\delta\sigma^2\rangle^{\kappa/2} 
\sigma_0^{1-\kappa}$ at small $\sigma_0$, with nonuniversal
dimensionless constant $C$ and universal critical exponent $\kappa$ characterizing the 
transport properties. While $\kappa=10/13$ is often quoted as an exact 
value~\cite{Isichenko,SH1994,PS1995,FS1995}, Simon and Halperin~~\cite{SH1994} argued 
that one could not completely rule out the possibility that finite dissipation may 
spoil the connection to geometrical percolation and change the value of 
$\kappa$. In order to check that this is not the case, we performed careful Pad\'e 
resummation~\cite{Supplementary} of the perturbative series~(\ref{eq:series}) up to 
six loops, see Table~\ref{TAT:tab:alpha}.
\begin{table}[htb]
\begin{tabular}{r|c|l}
Order & Method 	& Exponent $1-\kappa$ \\ 
\hline
\hline
2& Pad\'e 	& $0.28\pm0.09$\\ 
4& Pad\'e 	& $0.221\pm0.006$\\
4& n-fit 	& $0.233\pm0.002$\\
$\infty$&Conjecture& $3/13\simeq0.2308$
\end{tabular}
\caption{\label{TAT:tab:alpha} Critical exponent $1-\kappa$ obtained from
Pad\'e approximants~\cite{Supplementary} built from the perturbative
series~(\ref{eq:series}).}
\end{table}
Our most accurate result 
$\kappa=0.767\pm0.002$ seems to confirm the conjectured value
$\kappa=10/13\simeq0.7692$ based on the analogy to classical 
percolation~\cite{SH1994,PS1995,FS1995}. We stress the good 
convergence of the Pad\'e approximants for all values of the
dissipation strength $\sigma_0$, see Fig.~\ref{fig:extrapolation}.
Note that partial resummation of perturbation theory in previous works~\cite{Dreizin1972} 
failed to recover the critical behavior associated to percolation in the strong coupling 
regime and this approximation led to an incorrect saturation of $\sigma_{xx}$ in the limit
$\sigma_0\to0$, which would apply only for transport model with discrete conductivity
values~\cite{DR1994}.

\begin{figure}
 \includegraphics[width=0.8\linewidth]{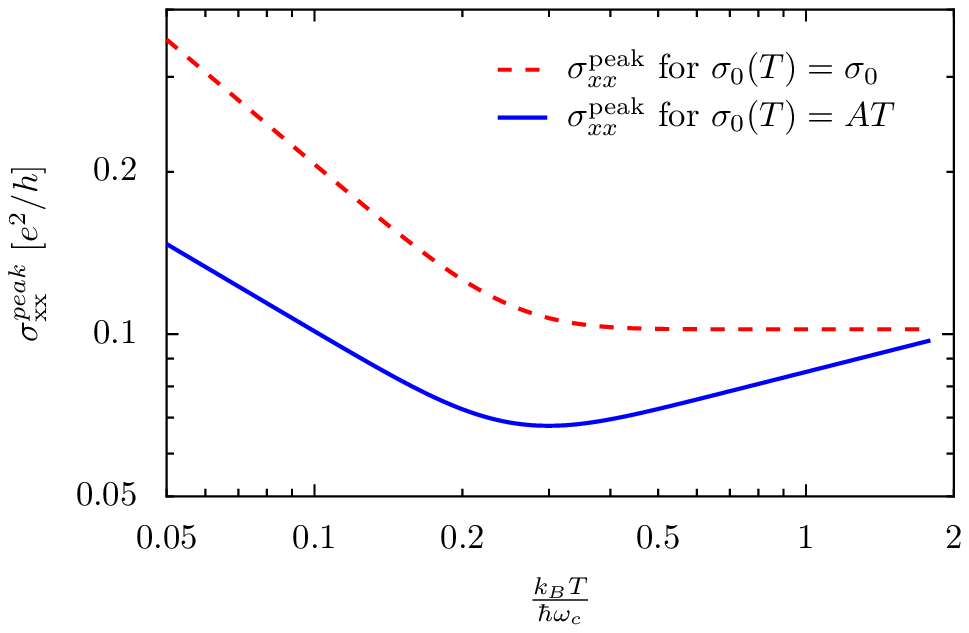}
 \caption{(color online) Temperature dependence of
the peak longitudinal conductance from~Eq.~(\ref{sigmapeak}) 
in log-scale.  A crossover occurs at
$k_B T^\star=\hbar\omega_c/4$ between a low-$T$ power-law
[$\sigma_{xx}\propto T^{-\kappa}$ or $T^{1-2\kappa}$] and a high-$T$
power-law [$\sigma_{xx}\propto \mathrm{cst}$ or $T^{1-\kappa}$], for
elastic scattering $\sigma_0(T)=\sigma_0$ or inelastic phonon
scattering $\sigma_0(T)=AT$ respectively.}
\label{Peak}
\end{figure}

{\it Microscopics of $\sigma_{xx}$ at plateau transitions.--} We finally study the temperature behavior 
of transport in the percolation dominated regime. At high magnetic field, the local 
Hall conductivity is explicitly related to the Fermi distribution of 
Landau levels $E_m=\hbar \omega_c (m+1/2)$ with integer $m$, disorder landscape 
$V(\ve{r})$ and chemical potential $\mu$~\cite{GV1994,CFC2008,Supplementary}:
\begin{equation}
\sigma_H(\ve{r}) = \frac{e^2}{h} \sum_{m=0}^{\infty} n_F[E_m+V(\ve{r})-\mu]
\label{sigmaFermi}
\end{equation}
neglecting spin effects.
We have introduced here the cyclotron energy $\hbar\omega_c=\hbar|eB|/m^\ast$ in terms of 
Planck's constant $\hbar$, electron charge $e$, applied perpendicular magnetic field $B$ and 
effective mass $m^\ast$.
At temperatures $T\gg\sqrt{\langle V^2\rangle}$, the Fermi
distribution $n_F(E)$ can be linearized, so that the random conductivity distribution~(\ref{TAT:eq:condTensor}) becomes Gaussian.
Straightforward analysis~\cite{Supplementary} and our low-dissipation
formula lead to a simple expression for the peak conductance measured at the
transition region between two Landau levels ($k_B$ is Boltzmann's constant):
\begin{equation}
\sigma_{xx}^\mathrm{peak}=\sigma_\mathrm{bg}(T,B)\left[1+\sum_{l=1}^\infty
\frac{4\pi^2l k_B T}{\hbar\omega_c}
\mathrm{csch}\left(\frac{2\pi^2l k_B T}{\hbar\omega_c}\right)\right]^\kappa\!\!.
\label{sigmapeak}
\end{equation}
This expression plotted in Fig.~\ref{Peak} shows 
a sharp 
crossover at temperature $k_B T^\star=\hbar\omega_c/4$ from a low-$T$ power-law 
behavior $\sigma_{xx}^{\mathrm{peak}}=\sigma_{\mathrm{bg}}(T,B)[\hbar\omega_c/(4k_BT)]^\kappa$~\cite{PS1995} 
to a high-$T$ background conductivity
\begin{equation}
\sigma_\mathrm{bg}(T,B)=C [\sigma_0(T,B)]^{1-\kappa} 
\left[\frac{e^2}{h}\frac{\sqrt{\langle
V^2\rangle}}{\hbar\omega_c}\right]^\kappa\!\!.
\label{sigmabg}
\end{equation}
%
Formulas~(\ref{sigmapeak})-(\ref{sigmabg}), which combine 
microscopic parameters (such as the width of the disorder distribution) with geometrical effects 
through the exponent $\kappa$, should be useful for detailed analysis of
transport measurement in quantum Hall samples.

Clearly, $\sigma_{xx}^{\mathrm{peak}}$ cannot diverge at $T\to0$ and is in fact expected to 
level off when reaching conductance values of the order of $e^2/2h$~\cite{DR1994,GammelEvers}.
In this very-low-temperature regime, the linearization of the local
Hall conductivity (\ref{sigmaFermi}) breaks down, thereby putting
a limit to the present diagrammatic calculation. 
Furthermore, quantum effects become important at low-$T$ and 
lead~\cite{Kettemann,Li2010} to a different exponent 
$\kappa^\mathrm{qu.}\simeq3/7\simeq0.42$. The classical percolation
exponent~\cite{SH1994,PS1995,FS1995} $\kappa=10/13\simeq0.77$ may 
be observable in very high mobility samples dominated by smooth 
disorder~\cite{Zhao2008,Li2010}. Finally, at temperatures 
$T>\hbar\omega_c/4$, the leading magnetic field dependence of the longitudinal
conductivity in Eq.~(\ref{sigmapeak}) is provided by the $\omega_c^{-\kappa}
\propto B^{-\kappa}$ term, as 
discussed  previously~\cite{PEMW201,Renard2004}. 

{\it Conclusion.--}
We have used a general diagrammatic method to compute fully microscopically
the electronic transport in incoherent disordered conductors, leading
to accurate determination of critical exponents for the conductivity in the 
classical percolation regime of the quantum Hall transition.
This framework seems also well suited for efficient numerical implementations using 
the recently developed diagrammatic Monte Carlo methods~\cite{Gull2011}, leading to envision 
progresses towards more realistic description of quantum Hall transport taking into 
account disorder effects.

\begin{acknowledgments}
We thank A. Freyn for precious help with symbolic computation, and
S. Bera, B. Piot, M. E. Raikh, V. Renard and F. Schoepfer 
for stimulating discussions.
\end{acknowledgments}


\newpage
\setcounter{figure}{0}
\setcounter{table}{0}
\setcounter{equation}{0}

\section{Supplementary Material for ``Diagrammatic Approach for the
High-Temperature Regime of Quantum Hall Transitions''}

\subsection{Evaluation of the diagrams}\label{TAT:app:diag}

We consider here the problem of random Gaussian fluctuations of the local 
Hall conductivity in two dimensions (see Eq.~(14) in the main text), split into 
an average Hall component $\sigma_H$ and a
fluctuating term $\delta\sigma(\ve{r})$, defined so that
$\left<\delta\hat\sigma(\ve{r})\right>=0$. The dissipationless nature
of the Hall component shows up by the fact that $\sigma_H$ exactly drops 
in the correlation function $\left[\hat{\cal{G}}_0\right]_{ij}
=\frac{\partial}{\partial r_i}\frac{\partial}{\partial r_j}G(\ve{r},\ve{r'})$:
\begin{align}
\left[\hat{\cal{G}}_0\right]_{ij}(\ve{r})
=-\frac{1}{\sigma_0}\int\!\!\frac{d^2p}{(2\pi)^2}
\frac{p_i p_j e^{i\ve{p}\cdot\ve{r}}}{\ve{p}^2+0^+},
\end{align}
with $G(\ve{r},\ve{r'})$ defined by Eq.~(11) in the main text.

The first order diagram contributing to the conductivity is straightforwardly
calculated in the case of Gaussian fluctuations of the Hall component
in two dimensions (see Eq.~(14) in the main text):
\begin{align}
\nonumber
&\includegraphics[bb=-3 -3 26 14]{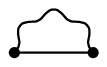} 
=\int\!\!\frac{d^2 \ve{p}}{(2\pi)^2}\tilde{K}(\ve{p})\hat{\epsilon}\hat{\cal{G}}_0(\ve{p})\hat{\epsilon}\\
\nonumber
&=\frac{\langle \delta\sigma^2\rangle}{\pi\sigma_0} \!\! \int_0^{+\infty} 
\!\!\!\!\!\!\!\!\! dp\,
pe^{-p^2} \!\!\!\! \int_0^{2\pi} \!\!\!\!\!\!\! d\theta
\begin{pmatrix} 
\sin^2(\theta)& -\cos(\theta)\sin(\theta)\\
-\cos(\theta)\sin(\theta) & \cos^2(\theta) \end{pmatrix}\\
&=\frac{\langle \delta\sigma^2\rangle}{2\sigma_0} 
\begin{pmatrix} 1& 0\\ 0 & 1 \end{pmatrix}
\label{OrderTwo}
\end{align}
with $K(\ve{r})\equiv \langle\delta\sigma(\ve{r})\delta\sigma(\ve{0})\rangle=
\langle\delta\sigma^2\rangle e^{-|\ve{r}|^2/\xi^2}$, and its
Fourier transform $\tilde{K}(\ve{p}) = \pi \xi^2 \langle \delta\sigma^2\rangle
e^{-\xi^2 \ve{p}^2/4}$. Here $\hat\epsilon$ denotes the fully antisymmetric
$2\times2$ matrix,
$\hat\epsilon=\big[\phantom{\big[}_1^0\phantom{\big[}_{\phantom{-}0}^{-1}\big]$.
Note that the conductivity correction [Eq.~(\ref{OrderTwo})] is positive and exactly opposite 
in sign to the one obtained in the case of pure longitudinal fluctuations of the
conductivity in Eq.~(13) of the main text.

All second and third order diagrams can be obtained analytically with
the help of symbolic computation, see the results displayed in
Table~\ref{D:tab:4thAnd6thOrder}.
\begin{table*}
\label{secondandthird}
\begin{tabular}{c|r|l|l}
Diagram & Multiplicity & Analytical Value & Decimal Value\\ 
\hline
\hline
second order& & &\\
\includegraphics[width=1.5cm]{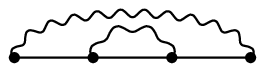}
&1&$-\frac{1}{4}\log(2)$ &-0.173287 \\
\includegraphics[width=1.5cm]{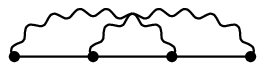}
&1&$\frac{1}{8} (1 - \log(4)) $&-0.0482868 \\
\hline
third order& & &\\
\includegraphics[width=2cm]{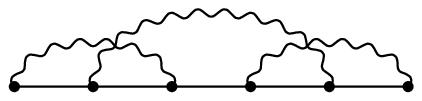}
 & 1 & $\frac{1}{96} \left(3 - \pi^2 + 3 \log[3] (-3 + \log[9]) 
+ 12 \mathrm{Polylog}\left[2, \frac{2}{3}\right]\right)$\normalsize & 0.00504001\\ 
\includegraphics[width=2cm]{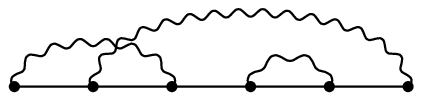}
& 2 & $\frac{1}{32} \log\left[\frac{27}{16}\right]$ & 0.0163515\\ 
\includegraphics[width=2cm]{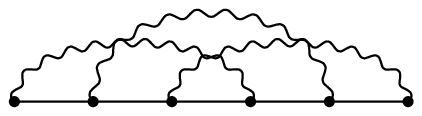}
 & 1 & $\frac{1}{16} \left(2 \log[2]^2 - 3 \log[3] + \log[8] 
+ \mathrm{Polylog}\left[2, \frac{1}{4}\right]\right)$\normalsize & 0.000760209\\ 
\includegraphics[width=2cm]{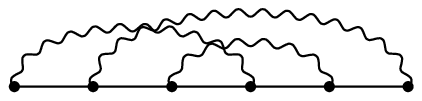}
 & 2 & $\frac{1}{384} (2 + 100 \log[2] - 63 \log[3])$ & 0.00547433\\ 
\includegraphics[width=2cm]{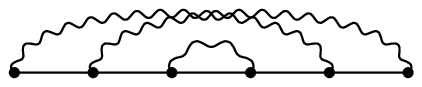}
 & 1 & $\frac{1}{8} \log\left[\frac{32}{27}\right]$ & 0.0212374\\ 
\includegraphics[width=2cm]{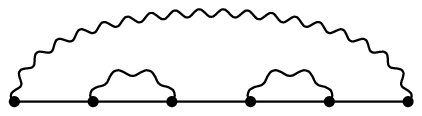}
& 1 & $\frac{1}{8} \log\left[\frac{27}{16}\right]$ & 0.065406\\ 
\includegraphics[width=2cm]{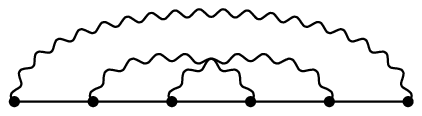}
 & 1 & $-\frac{1}{48} -\frac{ \log[2]}{6} + \frac{9 \log[3]}{64}$ & 0.0181345\\ 
\includegraphics[width=2cm]{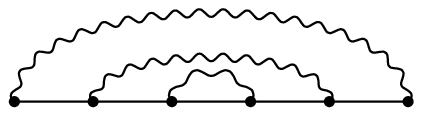}
 & 1 & $\frac{3}{16} \log\left[\frac{4}{3}\right]$ & 0.0539404
\end{tabular}
\caption{\label{D:tab:4thAnd6thOrder}Diagrams to second and third order: 
multiplicity and analytical values. The resulting coefficients $a_4$ and $a_6$
are given in Table~\ref{values}.}
\end{table*}
The method of computation for the second and third order contributions is
to first express each of the several denominators appearing in a given 
graph using Feynman's identity:
\begin{equation}
 \frac{1}{x_i}=\int_0^{\infty}\!\!\!\!\!\! dt_i \phantom{0} e^{-t_i x_i}.
\end{equation}
One can then perform the Gaussian integration over all momenta, and
finally compute the remaining integrals over the auxiliary variables $t_i$.

\begin{figure}
 \includegraphics[bb=-17 -12 109 35]{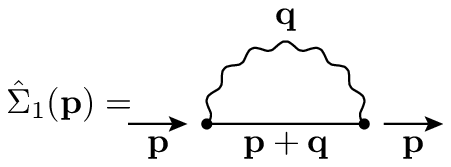}\\
\caption{\label{D:fig:SigmaP}Self-energy $\hat\Sigma_1(\ve{p})$ entering the calculation of 
the non-crossing diagrams in Table~\ref{TAT:tab:triumphal_arch_diagrams}.}
\end{figure}

We have not managed to analytically obtain the diagrams of fourth order and beyond
(except for the non-crossing ones, see below), and we had therefore recourse to a
combination of analytical and numerical steps. First, an automated script was
used to generate all possible diagrams, discarding the 
particle reducible ones, which enables to output explicitely the corresponding functions
that require full momentum integration. In order to avoid indefinite integrals, all
two-dimensional momenta in an $n^\mathrm{th}$ order diagram were combined into 
the hyperspherical coordinate $\ve{K}$ in dimension $2n$, such that 
$\ve{K}^2=\sum_{i=1}^{n} {\ve{p}_i}^2$. This allows analytical integration over
$|\ve{K}|$, leaving the bounded integration domain on the hypersphere in $2n$
dimensions. This numerical step was finally performed using the
\texttt{Vegas} Monte Carlo integration routine from the \texttt{GNU Scientific
Library}. Because only the complete sum of all diagrams at a given order matters,
and since multidimensional integrals are time consuming, we have 
summed up all the contributions at a given order before performing the integration.
The Monte Carlo evaluations were iterated until the relative error was below
$0.1\%$, but we can also ascertain the good convergence of the numerics by
benchmarking the routine on analytically tractable diagrams that have 
no crossings of the propagators, see Table~\ref{TAT:tab:triumphal_arch_diagrams}
for comparison.
\begin{table*}
\begin{tabular}{c|c|l|l}
Order & Diagram & Analytical value & Monte Carlo evaluation\\ 
\hline
\hline
4 & \includegraphics[width=2cm]{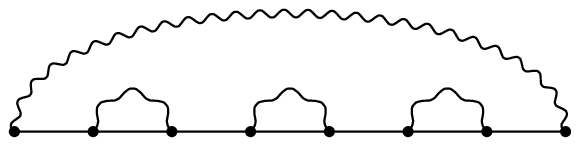} 
& $\frac{-44 \log[2] + 27 \log[3]}{32}\simeq$ -0.02612 & -0.02607\\ 
5 & \includegraphics[width=2.5cm]{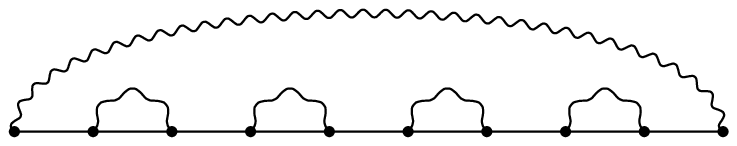}
& $\frac{162 \log[3] + 125 \log[5] - 544 \log[2]}{192}\simeq 0.01084$ & 0.01087 \\ 
6 & \includegraphics[width=3cm]{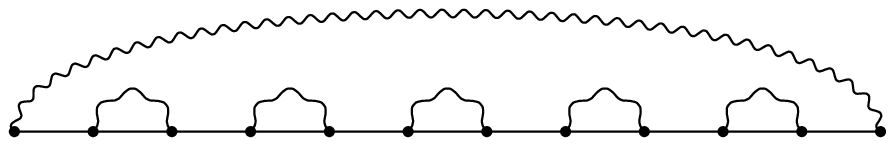}
& $\frac{-6496 \log[2] - 486 \log[3] + 3125 \log[5]}{1536}\simeq-0.004632$ & -0.004630
\end{tabular}
\caption{\label{TAT:tab:triumphal_arch_diagrams} Benchmarking the numerical Monte
Carlo evaluation against analytically tractable non-crossing diagrams at fourth,
fifth and sixth order respectively.}
\end{table*}
The high (up to 6$^\mathrm{th}$) order non-crossing diagrams that we considered
are obtained in the following way: we remark that
these graphs are only composed of bare propagators and of the first order
self-energy $\hat\Sigma_1$ appearing in Fig.~\ref{D:fig:SigmaP}. 
The momentum dependence of this self-energy is readily evaluated:
\begin{eqnarray}
\hat\Sigma_1(\ve{p})&=&\frac{\langle\delta\sigma^2\rangle}{\sigma_0}
\frac{1}{(p_x^2+p_y^2)^2} 
\begin{pmatrix}
a&b\\
b&c\\
\end{pmatrix},\\
\nonumber
a&=& \frac{1}{2}(p_y^2 - p_x^2) \left[e^{-p_x^2-p_y^2}-1\right]
+  p_x^2 p_y^2 + p_y^4,\\
\nonumber
b&=& -p_x p_y [e^{-p_x^2 - p_y^2}-1 + p_x^2 + p_y^2],\\
\nonumber
c&=& \frac{1}{2}(p_x^2 - p_y^2) \left[e^{-p_x^2-p_y^2}-1\right]
+ p_x^2 p_y^2 + p_x^4.
\end{eqnarray}
We note that $\hat\Sigma_1(\ve{p}=\ve{0})$ recovers the first order contribution to
the conductivity in Eq.~(\ref{OrderTwo}). At finite momentum, the self-energy
contains off-diagonal elements, although the final correction to the conductivity 
is purely diagonal.
The analytical computation of the non-crossing diagrams then proceeds as
previously described, using Feynman's trick and Gaussian integration. For
instance the following fourth order contribution 
\begin{align}
\nonumber
&\includegraphics[width=2cm]{ordnung8plat_8234567.eps}=\int\!\!\frac{d^2p}{(2\pi)^2}\tilde{K}(\ve{p})
\hat{\epsilon}\hat{\cal{G}}_0(\ve{p})[\hat\Sigma_1(\ve{p})\hat{\cal{G}}_0(\ve{p})]^3
\hat{\epsilon}
\end{align}
only involves a single momentum integration, which can then be performed 
analytically. Its value is given in Table~\ref{TAT:tab:triumphal_arch_diagrams}.

\subsection{Extrapolation to the weak dissipation regime}\label{TAT:app:pade}

We present here the methodology to obtain the extrapolated behaviour of
the effective diagonal conductivity in the limit $\sigma_0\to0$, starting from
the large-$\sigma_0$ expansion:
\begin{equation}
 \sigma_{xx}(\sigma_0)=
\sigma_0\left[1+\sum_{n=1}^\infty
a_{n}\frac{\langle\delta\sigma^2\rangle^n}{\sigma_0^{2n}}\right]
\label{eq:pertseries}
\end{equation}
with the first six coefficients $a_n$ given in Table~\ref{values}.
\begin{table}[ht]
\begin{tabular}{r|c|l}
Order & Method 	& Coefficient $a_n$\\
\hline
\hline
1& Analytical & $\frac{1}{2}$ \\
2& Analytical & $\frac{1}{8}-\frac{1}{2}\log(2)$\\
3& Analytical & $0.2034560502$\\
4& Numerical & $-0.265\pm0.001$\\
5& Numerical & $0.405\pm0.001$\\
6& Numerical & $-0.694\pm0.001$
\end{tabular}
\caption{\label{values} Coefficients $a_n$ of the perturbative
series~(\ref{eq:pertseries}) up to sixth loop order.}
\end{table}

One standard method of extrapolation is the so-called DLog Pad\'e approximants~\cite{Singh1987}, 
which starts with the dimensionless logarithmic derivative of the function to extrapolate:
\begin{equation}
f(x)\equiv \frac{\sigma_0}{\sigma_{xx}(\sigma_0)}
\frac{\mathrm{d}\sigma_{xx}(\sigma_0)}
{\mathrm{d} \sigma_0}_{\big|\sigma_0/\sqrt{\langle\delta\sigma^2\rangle}\to x}.
\label{fdef}
\end{equation}
One then reexpands at small $x$ the function $f(x)$ to order $N$:
\begin{equation}
f_N(x)= 1+\sum_{n=1}^N b_{n} x^{2n}
\label{eq:fseries}
\end{equation}
with the coefficients $b_n$ given in Table~\ref{tableb}.
\begin{table}[ht]
\begin{tabular}{c||c|c|c|c|c|c}
Order & 1 & 2 & 3 & 4 & 5 & 6 \\
\hline
Coefficient $b_n$ & -1 & $\log(4)$ & -2.135 & 3.698 & -6.919 & 13.823
\end{tabular}
\caption{\label{tableb} Coefficients $b_n$ used in the DLog Pad\'e
extrapolation, corresponding to the small-$x$ series expansion~(\ref{eq:fseries})
of the function $f(x)$ defined in Eq.~(\ref{fdef}).}
\end{table}
The DLog Pad\'e method uses then an approximant for $f(x)$ of the following form:
\begin{equation}
f_N(x) = \frac{1+\sum_{n=1}^N c_{n} x^{2n}}
{1+\sum_{p=1}^N d_{n} x^{2n}}.
\label{eq:dlogseries}
\end{equation}
The coefficients $c_n$ and $d_n$ are computed from the knowledge
of the perturbative terms $b_n$ given in Table~\ref{tableb}.
>From the expected power-law behavior of the conductivity at small dissipation,
$\sigma_{xx}\propto \langle\delta\sigma^2\rangle^{\kappa/2} 
\sigma_0^{1-\kappa}$, one gets $f(x)\to(1-\kappa)$ for $x\to\infty$. 
The critical exponent $\kappa$ is thus obtained by extrapolating the
Pad\'e approximant~(\ref{eq:dlogseries}) to infinity, which simply
reads $1-\kappa=c_{N}/b_{N}$ at the order $N$.

The corrections to the effective conductivity at second order require
an order $N=2$ DLog Pad\'e approximant, which lead after integration of
Eq.~(\ref{fdef}) to the formula:
\begin{equation}
\sigma_{xx}\simeq\sigma_0\left[1+\frac{1}{\kappa}
\frac{\langle\delta\sigma^2\rangle}{\sigma_0^2}\right]^{\kappa/2}
\label{sigmaApprox}
\end{equation}
with $\kappa=0.72\pm0.09$. The error bar on $\kappa$ is obtained here by
expanding Eq.~(\ref{sigmaApprox}) to third order with $\kappa$
arbitrary, and comparing the deviation from the resulting
coefficient with the exact $a_3$ value. 
Eq.~(\ref{sigmaApprox}) captures the full crossover between the 
perturbative regime $\langle\delta\sigma^2\rangle\ll \sigma_0^2$ (where strong 
dissipation controls transport) to the non-perturbative limit of vanishing 
dissipation $\sigma_0\to0$ (where percolation effects dominate), see 
Fig.~4 in the main text.
%

In order to obtain a better
estimate for the exponent, one must push the calculation of the
effective conductivity to fourth order. Following the same strategy, 
the order $N=4$ DLog Pad\'e approximant provides the estimate 
$\kappa=0.779\pm0.006$, and the resulting formula for the effective 
conductivity takes the form:
\begin{equation}
\label{TAT:eq:padeform}
 \sigma_{xx}(\sigma_0)=\sigma_0
\left(1+A\frac{\langle\delta\sigma^2\rangle}{\sigma_0^2}\right)^{B}
\left(1+C\frac{\langle\delta\sigma^2\rangle}{\sigma_0^2}\right)^{D}
\end{equation}
with dimensionless numbers $A, B, C, D$, leading to $\kappa=2B+2D$. 
Again, the error bar on $\kappa$ is obtained from
comparison to the next known coefficient, namely $a_{5}$, expanding 
Eq.~(\ref{TAT:eq:padeform}) to fifth order while keeping an arbitrary 
$\kappa$ fixed (a small additional error due to the Monte Carlo evaluation 
of the coefficients was also taken into account).

While our calculation of the sixth order corrections to the conductivity 
would allow us in principle to further refine the estimation of the
exponent, we encounter in that case a spurious pole~\cite{Watts1975}, that 
invalidates the method. One explanation why the Pad\'e method
becomes unstable at high orders can be understood already from
the fourth order extrapolation~(\ref{TAT:eq:padeform}), which leads to
trivial sub-leading corrections to scaling at small dissipation:
\begin{equation}
\label{TAT:eq:pade4expand}
 \sigma_{xx}(\sigma_0) \propto 
\langle\delta\sigma^2\rangle^{\kappa/2} 
\sigma_0^{1-\kappa}\,\left 
[1+E \frac{\sigma_0^2}{\langle\delta\sigma^2\rangle} +\dots \right].
\end{equation}
This shows that the DLog Pad\'e method enforces a given value
$\kappa'\simeq3-\kappa$ for the sub-leading exponent $\kappa'$, which is 
unlikely to correspond with good precision to the right value. This lack 
of flexibility is the likely source of the instability of the Pad\'e approximant, and 
authors~\cite{Ferer1983} have used a generalized n-Fit method
that circumvents this problem. For the case of the fourth order
conductivity, the fitting formula has rather the following additive
form:
\begin{equation}
\sigma_{xx}(\sigma_0)=F\sigma_0
\left(1+G\frac{\langle\delta\sigma^2\rangle}{\sigma_0^2}\right)^{H}
\!\!\!\!+(1-F)c_0\left(1+I\frac{\langle\delta\sigma^2\rangle}{\sigma_0^2}\right)^{J}\!\!\!\! .
\label{TAT:eq:fit}
\end{equation} 
The critical exponent is then given by $\kappa=\min[2H,2J]$,
while the independent subleading exponent reads $\kappa'=\max[2H,2J]$.
All unknown numerical coefficients are obtained by expanding
Eq.~(\ref{TAT:eq:fit}) at small $x$ and fitting to the coefficients of
Table~\ref{tableb}.
Estimating the error by comparison to the known $a_{5}$ coefficient,
we find $\kappa=0.767\pm0.002$, in excellent agreement with the
conjectured value $\kappa=10/13\simeq0.7692$. 
Moreover, the Pad\'e approximants show good convergence for all values of the
dissipation strength $\sigma_0$, see Fig.~4 in the main text.

\subsection{High temperature microscopics of ${\bm \sigma_{xx}}$ at the plateau transition}

The local Hall conductivity can be computed microscopically in the high
magnetic field regime~\cite{supGV1994,supCFC2008}, and simply follows from Hall's law 
with Landau level quantization:
\begin{equation}
\sigma_H(\ve{r}) = \frac{e^2}{h} \sum_{m=0}^{\infty} n_F[E_m+V(\ve{r})-\mu]
\label{sigmaHighfield}
\end{equation}
with standard Landau levels $E_m=\hbar \omega_c (m+1/2)$, cyclotron frequency 
$\omega_c=|eB|/m^\ast$, random disorder potential 
$V(\ve{r})$, chemical potential $\mu$ and Fermi function $n_F(E)=1/\{\exp[E/(k_B T)]+1\}$. 
In particular, microscopic calculations~\cite{supCFC2008} show that deviations to the
form~(\ref{sigmaHighfield}) are small by the dimensionless parameter 
$[l_B^2 \sqrt{\langle V^2\rangle}]/[\xi^2 \hbar \omega_c]\ll1$, with 
$\sqrt{\langle V^2\rangle}$ the width of the disorder distribution,
$l_B=\sqrt{h/|eB|}$ the magnetic length, and $\xi$ the large correlation length of 
the disorder fluctuations. Note the smallness of $l_B\simeq8$nm at $B=10$T, so
that $l_B\ll\xi$ for smooth disorder.

At temperatures such that $T\gg\sqrt{\langle V^2\rangle}$, the Fermi
distribution in Eq.~(\ref{sigmaHighfield}) can be linearized, so that Gaussian 
fluctuations of disorder provide Gaussian fluctuations for the Hall
conductivity $\sigma_H(\ve{r})=\sigma_H + \delta\sigma(\ve{r})$ with
\begin{eqnarray}
\sigma_H &=& \frac{e^2}{h} \sum_{m=0}^{\infty} 
n_F(E_m-\mu),\\
\delta\sigma(\ve{r})&=& \frac{e^2}{h} \sum_{m=0}^{\infty} 
n_F'(E_m-\mu) V(\ve{r}).
\label{sigmaFluct}
\end{eqnarray}
The power-law behavior of the longitudinal conductivity at small
dissipation, $\sigma_{xx}= C \sigma_0^{1-\kappa} 
\langle\delta\sigma^2\rangle^{\kappa/2}$, leads to:
\begin{equation}
\sigma_{xx}=C \sigma_0^{1-\kappa} 
\left|\frac{e^2}{h}\sqrt{\langle V^2\rangle}
\sum_{m=0}^{\infty} n_F'(E_m-\mu) \right|^\kappa\!.
\label{sigmaAlmost}
\end{equation}
We re-express the sum over Landau levels in Eq.~(\ref{sigmaAlmost}) by using Poisson
summation formula:
\begin{equation}
\sum_{m=0}^{+\infty} f(m)=\sum_{l=-\infty}^{+\infty} \int_0^{+\infty} \!\!\!\!\!  dt\,
e^{i 2\pi l t} f(t).
\end{equation}
In the limit $T<\mu$, one finds after standard manipulations~\cite{CM2001}:
\begin{eqnarray}
\nonumber
\left|\sum_{m=0}^{\infty} n_F'(E_m-\mu)\right|&=&
\frac{1}{\hbar \omega_c}\Bigg[1+\sum_{l=1}^{+\infty}
(-1)^l \cos\left(\frac{2\pi l\mu}{\hbar\omega_c}\right) \\
&& \times \frac{\frac{4\pi^2 l k_B T}{\hbar\omega_c}}
{\mathrm{sinh}\left(\frac{2\pi^2 l k_B T}{\hbar\omega_c}\right)}\Bigg].
\end{eqnarray}
Finally, by considering the plateau transition region between the 
filling factors $\nu$ and $\nu+1$, the chemical potential is
pinned to $\mu=\hbar\omega_c(\nu+1/2)$, leading to expressions (17)-(18) 
in the main text.

\end{document}